\pdfoutput=1 
\documentclass[
reprint,
showpacs,
preprintnumbers,
superscriptaddress,
amsmath,amssymb,
aps,
]{revtex4-1}

\usepackage{amsmath}
\usepackage{amssymb}
\usepackage{mathtools,amssymb}
\usepackage{amsthm}
\usepackage[utf8]{inputenc}
\usepackage{graphicx}
\usepackage{dcolumn}
\usepackage{bm}
\usepackage{hyperref}
\usepackage{url}

\newcommand{\dg}{\dagger}
\newcommand{\e}{\mathcal{E}}
\renewcommand{\oe}{\overline{\mathcal{E}}}

\newcommand{\tr}{\mathrm{tr}}
\newcommand{\id}{\mathbb{I}}
\renewcommand{\S}{\mathcal{S}}
\newcommand{\mb}{\mathbf}
\renewcommand{\sp}{\mathrm{span}}
\renewcommand{\P}{\mathbb{P}}

\newcounter{app}

\begin{document}


\title{Matrix Product States with long-range Localizable Entanglement}

\author{T. B. Wahl}\email{thorsten.wahl@mpq.mpg.de}
\affiliation{Max-Planck Institut f\"{u}r Quantenoptik, Hans-Kopfermann-Str. 1, Garching, D-85748, Germany}

\author{D. P\'{e}rez-Garc\'{i}a}\affiliation{Departamento de Analisis Matematico. Universidad Complutense de Madrid, 28040 Madrid, Spain}

\author{J. I. Cirac}
\affiliation{Max-Planck Institut f\"{u}r Quantenoptik, Hans-Kopfermann-Str. 1, Garching, D-85748, Germany}

\date{\today}

\begin{abstract}
We derive a criterion to determine when a translationally invariant matrix product state (MPS) has long-range localizable entanglement, where
that quantity remains finite in the thermodynamic limit. We give examples fulfilling this criterion and eventually use it to obtain all such MPS with bond dimension 2 and 3. 
\end{abstract}

\maketitle


Localizable entanglement (LE) \cite{LE} is a multipartite measure characterizing the (maximal averaged) bipartite entanglement present in a system. Motivated by quantum repeaters \cite{repeaters}, where it provides the natural figure of merit, the LE has also been applied to many-body physics problems \citep{LEMPS,LECondMat, Gaussian, Ent_trans, 
Sahoo}, whereby it can reveal hidden correlations, not detectable by standard observables \cite{LEMPS}. 

In both scenarios, there is an underlying spatial structure. Hence, peculiarities are expected whenever a finite amount of entanglement between two particles with arbitrary distance can be created by properly measuring the rest of the particles. Such states are said to have long-range localizable entanglement (LRLE) \cite{Ent_Length} and play an important role 
both in the context of quantum repeaters and
spin chains. In the first one, they are those for which entanglement can be established at arbitrary distances. In the latter, phase transitions 
are signaled by a finite value of the LE between any two particles irrespective of their distance. 
States with LRLE thus play a crucial role in those two contexts. But, which are those states?

As most multipartite entanglement measures, the LE is very hard to determine for general states. A notable exception is the set of matrix product states (MPS) \cite{MPS}. This family of states describes the ground state of gapped 1D spin
chains as well as those states created by sequential generation, as it is the case of atoms in a cavity \cite{SchoenPRL}. They are characterized by a
set of rank-three tensors, each one associated to a spin. One of the indices
corresponds to the spin in the $z$-basis, and the other two run from $1$ to $D$, where $D$ is called the bond dimension. MPS of arbitrary bond dimension
are dense in the set of all multipartite states \cite{FNW92}, and thus, they are very
relevant to describe many-body systems. For translationally invariant systems and low bond dimensions ($D=2$) the LE can be determined exactly
 \cite{LEMPS}. For larger bond dimensions, one may find relatively tight lower
bounds using Monte-Carlo Methods \cite{LE_long, QMC}.

In this paper we fully characterize translationally invariant MPS with LRLE for arbitrary bond dimension. In particular, we give a set of necessary and sufficient conditions for such states. As we show, those conditions can be turned into a set of polynomial equations, and thus provide us with a precise criterion to determine if a state has LRLE or not. Furthermore, we give examples of non-trivial states with that property and provide the full sets of MPS with LRLE for bond dimension $D = 2$ and $D = 3$.



The LE is defined as the maximum average entanglement that can be generated between two spins of a spin chain by measuring the remaining ones \cite{LE}. 
Let $\rho$ denote the density matrix of the original state. With probability $p_{\mb i}$ the outcome of a measurement $\mathcal{M}$ will be $\mb{i}$ and the system will be in the corresponding two-particle state $\rho_{\mb i}$. Hence, the LE is given by
\begin{align}
L^{\mathcal{C},E}(\rho) = \sup_{\mathcal{M} \in \mathcal{C}} \sum_{\mb{i}} p^\mathcal{M}_{\mb{i}} E(\rho^\mathcal{M}_{\mb i}),
\end{align}
where $\mathcal{C}$ is the class of allowed measurements and $E(\cdot)$ an entanglement measure.


Our system of consideration is an open chain of $N$ spin-$S$ particles along with two auxiliary particles of spin $S'$ at each of the boarders. The $N$ particles of the actual chain are the ones to be measured (measurement outcomes $\mb i = (i_1, \ ..., \ i_N)$), and the class of allowed measurements $\mathcal{C}$ is the set of local projective von Neumann measurements, where the same measurement is carried out on each party (in particular, we exclude adaptive strategies). Therefore, the maximization of the average entanglement is performed by choosing the optimal physical basis $\{| i \rangle \}_{i=1}^{2S+1}$.

\begin{figure}
\centering\includegraphics[scale=.14]{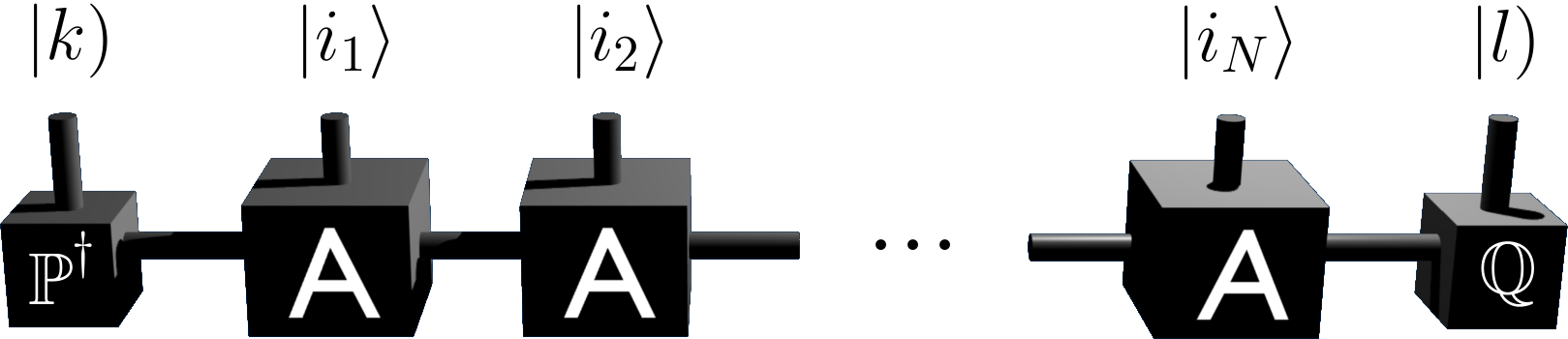}
\caption{Spin chain with $N$ spin-$S$ particles and one auxiliary particle of spin $S'$ at each of the borders. The matrices of the real particles are $A_i$ and those of the auxiliary particles $\P^\dg$ and $\mathbb{Q}$, respectively.}
\label{fig}
\end{figure}

The question to be answered in this letter is for which translationally invariant MPS a finite amount of entanglement can be localized between the two ancillas in the limit $N \rightarrow \infty$. We assume the state of the system to be translationally invariant apart from boundary effects; for this reason the rank-three tensors corresponding to the spin-$S$ particles are taken equal. Those consist of complex $D \times D$ matrices $A_i$ ($i = 1, \ ..., \ d \equiv 2S + 1$) and can be taken to be in canonical form, in which the maps $\e(X) = \sum_{i=1}^d A_i X A_i^\dg$ and $\oe (X) = \sum_{i=1}^d A_i^\dg X A_i$ satisfy (cf. \cite{MPS})
\begin{align}
\e(\id) = \id, \  \  \  \oe(\Lambda) = \Lambda
\label{gauge}
\end{align}
for some diagonal positive definite matrix $\Lambda$. Our goal is to find necessary and sufficient conditions on the matrices $\{A_i\}_{i=1}^d$ to give rise to LRLE for some matrices of the auxiliary particles. Those can be chosen at will and are denoted by $\P,\mathbb{Q}: \mathbb{C}^{D'} \rightarrow  \mathbb{C}^D$, where $D' = 2S' + 1 \leq D$ is the Hilbert space dimension of the individual auxiliary spins. The initial MPS is therefore
\begin{align}
| \psi \rangle = \sum_{k,l=1}^{D'} \sum_{i_1, ..., i_N = 1}^d (k|\P^\dg A_{i_1} ... A_{i_N} \mathbb{Q}|l) | i_1 ... i_N \rangle \otimes |k,l),
\end{align}
where the Hilbert space vectors of the auxiliary particles are denoted by round brackets, c.f. Fig. \ref{fig}.
Subsequent to a measurement $\mathcal{M}$ the initial state of the system $\rho = | \psi \rangle \langle \psi |$ reduces with probability $p_{\mb i} = \tr(\P^\dg A_{i_1} ... A_{i_N} \mathbb{Q} \mathbb{Q}^\dg A_{i_N}^\dg ... A_{i_1}^\dg \P)$ to $\rho^\mathcal{M}_{\mb i} = | \psi^\mathcal{M}_{\mb i} \rangle \langle \psi^\mathcal{M}_{\mb i}|$, where (excluding cases with $p_{\mathbf i} = 0$)
\begin{align}
| \psi^\mathcal{M}_{\mb i} \rangle = \frac{1}{\sqrt{p_{\mb i}}} \sum_{k,l=1}^{D'} (k|\P^\dg A_{i_1} ... A_{i_N} \mathbb{Q}|l)|kl)
\end{align}
is the normalized two-particle state after the measurement.
In Appendix \ref{app_show_qubits} it is shown that any MPS with LRLE for $D' > 2$ also has LRLE for $D' = 2$ (the converse is obvious), and also that w.l.o.g. we can take $\P$ and $\mathbb{Q}$ as isometries. We can thus choose $D' = 2$ and the concurrence \cite{concurrence} as the measure of entanglement, $E(\rho_{\mb i}^\mathcal{M}) = 2|\det ({\Psi}_{\mb i})|$, where ${\Psi}_{\mb i}$ is the matrix with coefficients $(\Psi_{\mb i})_{kl} = \frac{1}{\sqrt{p_{\mb i}}} (k|\P^\dg A_{i_1} ... A_{i_N} \mathbb{Q}|l)$.  Dropping the superscripts $\mathcal{C}$ and $E$, the LE reads
\begin{align}
L(\rho) = 2 \sup_{\{|i\rangle\}} \sum_{i_1, ..., i_N} \left| \det(\P^\dg A_{i_1} ... A_{i_N} \mathbb{Q}) \right|. \label{LE_det}
\end{align}
This optimization problem is in general hard, since the sum needs to be evaluated for large $N$ before being able to optimize over the physical basis $\{|i\rangle\}_{i=1}^d$. Its maximization succeeded only in special cases, like
%
for a chain of
$N$ spins without auxiliary ones and measurement on all spins but those at the borders \cite{AKLT}. In this case $\prod_{j=2}^{N-1} \sum_{i_j=1}^d | \det (A_{i_j}) |^{2/D}=1$, which implies that there is LRLE for $A_i = \alpha_i U_i$. In the following we slightly generalize this family of states and adapt them to our system that includes the auxiliary spins at the boundary:

\textit{Example 1.---Block structure of unitaries.}
\begin{align}
A_i = (P_i \otimes \id_{n \times n})\,  \bigoplus_{k=1}^{q} \, \alpha^k_i U^k_i,
\end{align}
where $P_i$ is a $q \times q$ permutation matrix. The $U^k_i$ are $n \times n$ unitaries ($D= n q$) and $\alpha^k_i \geq 0 \ \forall \ k = 1, \ ..., \ q, \ i = 1, \ ..., \ d$. 
We can realize that this MPS has LRLE for $n = 2$ by choosing $\P^\dg = |\uparrow)(1|$ + $|\downarrow)(2|$ (where $|\uparrow), \ |\downarrow)$ are basis vectors of the auxiliary qubits and $|1), \ |2), \ ...$ the basis vectors of the basis in which the matrices are given) and $\mathbb{Q} = D^{-1/2} \sum_{j=1}^D \left(|j)(\uparrow| + (-1)^j |j)(\downarrow| \right)$. Then, due to $\mathcal{E}(\mathbb{I}) = \mathbb{I}$ and therefore $\sum_{i=1}^d |\alpha_i^k|^2 = 1 \ \forall \ k = 1, \ ..., \ D/2$ \eqref{LE_det} is finite in the thermodynamic limit. The fact that there is also LRLE for $n > 2$ will become clear below. 

In the above examples the matrices $A_i$ exhibit a block structure of unitaries. One may think that all states with LRLE need to have this property in some basis $\{|i\rangle\}_{i=1}^d$.
Interestingly, this is not the case as shown by the following counterexample for $D = 3$:

\textit{Example 2.---Non-unitary matrices.}
\begin{align}
A_{1} = \frac{1}{2}\left(\begin{array}{ccc}
1&0&0\\
0&1&0\\
1&0&0
\end{array}\right)&, \
A_{2} = \frac{1}{\sqrt{2}}\left(\begin{array}{ccc}
0&0&1\\
0&1&0\\
0&0&-1
\end{array}\right), \notag \\
A_{3} =&\frac{1}{2}\left(\begin{array}{ccc}
0&1&0\\
1&0&0\\
0&1&0
\end{array}\right)
\end{align}
with $\P^\dg = |\uparrow)(1| + |\downarrow)(2|$ and $\mathbb{Q} = \frac{1}{\sqrt{3}} \left[ |1) + |2) + |3)\right] (\uparrow|$  $+ \frac{1}{\sqrt{2}}\left[|2) - |3)\right](\downarrow|$. It is a simple exercise to show that \eqref{LE_det} remains finite in the limit $N \rightarrow \infty$. The matrices $A_i$ fulfill $\e(\id) = \id$. However, $\oe(\mathbb{I}) \ne \mathbb{I}$, and thus they are not of the form $A_i = \alpha_i U_i$ in any basis $\{|i\rangle\}_{i=1}^d$. Note that the MPS can numerically be verified to be injective \cite{MPS}, and thus it is the unique ground state of a local translationally invariant gapped Hamiltonian. 
Moreover, it can be easily shown (analytically) that it is invariant under
the $Z_2$-symmetry generated by the transformation $|1\rangle \rightarrow |1\rangle$, $|2\rangle \rightarrow |2\rangle$, $|3\rangle \rightarrow - |3\rangle$. 



The question arises of how one can check whether a given MPS has LRLE without having to resort to evaluating \eqref{LE_det} for large $N$ numerically. 
In the following a necessary and sufficient criterion will be derived, which allows to decide this based on the matrices $\{A_i\}_{i=1}^d$ directly. 
We first rewrite \eqref{LE_det} by inserting to the right of $A_{i_j}$ a projector on the subspace spanned by the row vectors of $\P^\dg A_{i_1} ... A_{i_j}$, which can be written as $\P^{j}_{i_1 ... i_j} \P^{j \, \dg}_{i_1 ... i_j}$, where $\P^j_{i_1 ... i_j}: \mathbb{C}^2 \rightarrow \mathbb{C}^D$ is an isometry. 
Thus, \eqref{LE_det} reads now
\begin{align}
L(\rho) &= 2 \sup_{\{|i\rangle\}} \sum_{i_1} \left| \det (\P^\dagger A_{i_1} \P^1_{i_1})\right|
\sum_{i_2} \left| \mathrm{det} (\P^{1 \, \dagger}_{i_1} A_{i_2} \P^2_{i_1,i_2})\right| \notag \\
&\times ... \sum_{i_N} \left| \mathrm{det}(\P^{N-1 \ \dagger}_{i_1, ..., i_{N-1}} A_{i_N} \mathbb{Q}) \right|.
\label{projectors}
\end{align}
Using the SVD of $\P^{j-1 \dg}_{i_1,...,i_{j-1}} A_{i_j} \P^j_{i_1,..., i_{j}}$, which is $U_{i_j} \mathrm{diag}(\sigma_{i_j}^{(1)}, \sigma_{i_j}^{(2)}) V_{i_j}^\dg$ (neglecting the indices $i_1, \ ..., \ i_{j-1}$), along with the inequality of arithmetic and geometric means, one obtains for the factors of \eqref{projectors} 
\begin{align}
&\sum_{i_j} \left| \det (\P^{j-1 \dg}_{i_1,...,i_{j-1}} A_{i_j} \P^j_{i_1,..., i_{j}}) \right| \notag \\ &\leq \frac{1}{2} \sum_{i_j} \tr \left(\P^{j-1 \,\dg}_{i_1, ..., i_{j-1}} A_{i_j} \P^{j}_{i_1, ..., i_j} \P^{j \dg}_{i_1, ..., i_j} A^\dg_{i_j} \P^{j-1}_{i_1, ..., i_{j-1}} \right) \notag \\
&= 1,
\end{align}
since $\sum_{i_j} A_{i_j} A_{i_j}^\dg = \id$. The inequality becomes an equality if and only if $\sigma_{i_j}^{(1)} = \sigma_{i_j}^{(2)}$, i.e., $\P^{j-1 \dg}_{i_1,...,i_{j-1}} A_{i_j} \P^j_{i_1,..., i_{j}}$ is proportional to a unitary. The intuitive fact that all factors but the last of \eqref{projectors} have to be exactly 1 to get LRLE is proven in Appendix \ref{Lemma2}.
Therefore, a necessary condition for any MPS to give rise to LRLE is that there exists an isometry $\P^\dg$, such that for a certain basis $\{ |i \rangle \}_{i=1}^d$ 
\begin{align}
\P^{j-1 \dg}_{i_1,...,i_{j-1}} A_{i_j} \P^j_{i_1,..., i_{j}} \propto U_{i_1 ... i_{j}} \ \forall \ i_1, \ ..., \ i_j = 1, \ ..., \ d,
\label{prop_unitary}
\end{align}
$U_{i_1 ... i_{j}}$ denoting some $2 \times 2$ unitary. Redefining $\P^j_{i_1,..., i_{j}} \ \rightarrow \ \P^j_{i_1,..., i_{j}} U_{i_1 ... i_{j}}$ shows that one can require the RHS of \eqref{prop_unitary} to be the identity. After multiplying this from the right by $\P^{j \, \dg}_{i_1,..., i_{j}}$, one obtains
\begin{align}
\P^{j-1 \dg}_{i_1,...,i_{j-1}} A_{i_j} \P^j_{i_1,..., i_{j}} \P^{j \, \dg}_{i_1,..., i_{j}}
= \P^{j-1 \dg}_{i_1,...,i_{j-1}} A_{i_j} \propto \P^{j \, \dg}_{i_1,..., i_{j}},
\label{new_isometry}
\end{align}
which can also be written as
\begin{align}
\P^\dg A_{i_1} ... A_{i_j} \propto \P^{j \, \dg}_{i_1, ..., i_j}.
\label{basic_rel}
\end{align}
As illustrated by Example 1 and 2, $\mathbb{Q}$ can always been chosen such that \eqref{projectors} is finite, which is shown rigorously in Appendix \ref{app_suffQ}. 
Criterion \eqref{basic_rel} is thus necessary and sufficient for the emergence of LRLE \cite{D_eq2}.


If we define $\P^\dg := |\uparrow)(x| + |\downarrow)(y|$, i.e., $(x|x) = (y|y) = 1$ and $(y|x) = 0$, \eqref{basic_rel} is equivalent to
\begin{align}
(x| A_{i_1} ... A_{i_j} A_{i_j}^\dg ... A_{i_1}^\dg | x) &= (y | A_{i_1} ... A_{i_j} A_{i_j}^\dg ... A_{i_1}^\dg | y), \label{norm} \\
(y| A_{i_1} ... A_{i_j} A_{i_j}^\dg ... A_{i_1}^\dg | x) &= 0. \label{orth}
\end{align}
After further defining $V := |x)(x| - |y)(y|$, $W := |x)(y|$, the map $\oe_i (X) := A_i^\dg X A_i$ and
\begin{align}
V_{i_1, ... ,i_j} &:= A_{i_j}^\dg ... A_{i_1}^\dg |x)(x| A_{i_1} ... A_{i_j} \notag \\
& \ \ \ \ \ \ \ \ \ \ \ \ \ \ - A_{i_j}^\dg ... A_{i_1}^\dg |y) (y| A_{i_1} ... A_{i_j} \\
&= \oe_{i_j} \circ ... \circ \oe_{i_1} (V), \ \label{norm2} \\
W_{i_1, ... ,i_j} &:= A_{i_j}^\dg ... A_{i_1}^\dg |x)(y| A_{i_1} ... A_{i_j} \\
&= \oe_{i_j} \circ ... \circ \oe_{i_1} (W), \ \label{orth2}
\end{align}
we see that \eqref{norm} and \eqref{orth} are equivalent to
\begin{align}
\tr(V_{i_1, ... ,i_j}) &= \tr(\oe_{i_j} \circ ... \circ \oe_{i_1} (V)) = 0, \label{norm3} \\
\tr(W_{i_1, ... ,i_j}) &= \tr(\oe_{i_j} \circ ... \circ \oe_{i_1} (W)) = 0, \label{orth3}
\end{align}
respectively. Thus, if we define the subspace $\S := \mathrm{span} \{V, \ W,  \ ..., \ V_{i_1, ..., i_j}, \ W_{i_1, ..., i_j}, \ ... \}_{i_1, ..., i_j = 1}^d$, \eqref{norm3} and \eqref{orth3} indicate that the occurrence of LRLE is equivalent to $\tr(S) = 0 \ \forall \ S \in \S$. By definition $\S$ is closed under the application of any $\oe_i$, 
which leads to the following criterion characterizing translationally invariant MPS with LRLE:

\textit{Theorem 1.---For the MPS considered here, there is LRLE if and only if for a certain basis $\{|i\rangle\}_{i=1}^d$ there exists a subspace $\S$ of the vector space of $D \times D$ matrices satisfying the following conditions}
\begin{enumerate}
\item $\S$ \textit{is closed under all linear maps $\oe_i$ defined as $\oe_i(X) = A_i^{\dg} X A_i$, i.e., $\oe_i(\S) \subseteq \S \ \forall \ i = 1, \ ..., \ d$},
\item $\tr(S) = 0 \ \forall \ S \in \S$,
\item $\exists \ |x), \ |y) \ \in \mathbb{C}^D$ \textit{s.t.} $V = |x) (x| - |y)(y|$, $W = |x)(y|\in \S$.
\label{VW}
\end{enumerate}

Note that if Theorem 1 is fulfilled, $\oe_{i_j} \circ ... \circ \oe_{i_1} (W^\dg) = W_{i_1, ..., i_j}^\dg$ is also traceless, i.e., $\S$ could additionally be required to contain $W^\dg = |y)(x|$ and correspondingly to be equal to its adjoint, $\forall \ S \in \S \Rightarrow S^\dg \in \S$. 

Theorem 1 can be used numerically to determine, whether a given MPS has LRLE, 
since it imposes conditions on the matrices $\{A_i\}_{i=1}^d$, which can be represented by a set of polynomial equations: 
The entries of $\{A_i\}_{i=1}^d$ will give rise to the coefficients of those equations, whereas all other quantities introduced below will constitute their variables to be determined numerically. The first variables to be introduced are orthonormal basis vectors $\{S^k\}_{k=1}^n$ of $\S$ , where $n \leq D^2 - 1$ is the dimension of $\S$. Condition i is equivalent to requiring that any basis vector $S^k$ of $\S$ is mapped by any $\oe_i$ into $\S$, i.e., for all $k = 1, \ ..., \  n$
\begin{align}
\mathrm{I.}& \ \ \tilde A_i^\dg \, S^{k} \tilde A_i = \sum_{l=1}^n a_i^{k,l} S^l \ \forall \ i = 1, \ ..., \ d, \label{Groebner_first}\\
& \ \ \tr(S^{k \dg} S^l) = \delta_{k,l} \ \forall \ l = 1, \ ..., \ n, \\
\intertext{where $\{a_i^{k,l}\}_{i=1, ..., d}^{k,  l = 1,  ...,  n}$ are scalar complex variables of the set of equations to be solved numerically. The $\{\tilde{A_i}\}_{i=1}^d$ are the matrices in a possibly different physical basis $\{|\tilde i\rangle\}_{i=1}^d$. 
ii can be stated as}
\mathrm{II.}& \ \ \tr(S^k) = 0 \ \forall \ k = 1, \ ..., \ n. \\
\intertext{Furthermore, since according to condition iii $V = |x) (x| - |y)(y|$ and $W = |x)(y|$ have to be also contained in $\S$,}
\mathrm{III.}& \ \ \sum_{k=1}^n v^k S^k = |x)(x| - |y)(y|, \\
& \ \ \sum_{k=1}^n w^k S^k = |x)(y| \\
\intertext{with $\{v^k\}_{k=1}^n$, $\{w^k\}_{k=1}^n$, and the coefficients of $|x)$ and $|y)$ as other scalar complex variables. Last, a rotation in the basis of measurement $\{|i\rangle\}_{i=1}^d$ is implemented by}
\mathrm{IV.}& \ \ \tilde A_i = \sum_{j=1}^d U_{ij} A_j, 
\ \ U U^\dg = \id_{d \times d}, \label{Groebner_last}
\end{align}
constituting the last of the set of equations \eqref{Groebner_first} to \eqref{Groebner_last} to be solved. Generally, a set $p_i(x_1, \, ..., \ x_m) = 0$ of $s$ polynomial equations ($i = 1, \ ..., \ s$) with $m$ variables $\{x_j\}_{j=1}^m$ can be solved by means of a Gröbner basis \cite{Groebner}. A Gröbner basis $\{g_i(x_1, \ ..., \ x_m)\}_{i=1}^s$ is a special basis in the vector space of functions involving the variables $\{x_j\}_{j=1}^m$: It has the property that the set of equations to be solved is equivalent to the set $g_i(x_1, \, ..., \ x_m) = 0$ ($i = 1, \ ..., \ s$), which, in contrast, can be solved by back-substitution while having to deal with the solution of polynomials involving only one variable at a time. For instance, one of the new equations might involve only $x_1$. After solving it numerically the result can be inserted into another equation involving, e.g., only $x_1$ and $x_2$ etc. (cf. Gaussian elimination). A Gröbner basis can be found systematically by use of Buchberger's algorithm \cite{Groebner}, which is doubly exponential in the complexity of the set of equations to be solved. In our case, this implies a computational cost that is doubly exponential in the square of the bond dimension. However, $D$ is a constant and in particular independent of the length of the spin chain. If a simultaneous solution to \eqref{Groebner_first} to \eqref{Groebner_last} is found, the MPS has LRLE. If even for $n = D^2 - 1$ no solution is found, it does not.


We now employ Theorem 1 analytically to determine the complete sets of MPS with LRLE for $D = 2$ and $D = 3$. In the former case, we will reproduce the finding that the matrices need to be proportional to unitaries. In the latter, we obtain the result that matrices of the type of Example 2 are the only non-trivial matrices 
which give rise to LRLE. 

In both cases we take $\S$ to be equal to its adjoint, i.e., it has to contain the matrices $V = |x)(x| - |y)(y|$, $W = |x)(y|$ and $W^\dg$.
For $D = 2$ this implies that $\S$ is the full subspace of traceless $2 \times 2$ matrices. 
If we take any element $S \in \mathcal{S}$, $\tr (A_i^\dg S A_i) = 0$ shows that $A_i A_i^\dg$ is orthogonal to $\mathcal{S}$ with respect to the Hilbert-Schmidt scalar product. However, the orthogonal subspace of $\mathcal{S}$ is spanned by the identity, i.e., $A_i A_i^\dg \propto \mathbb{I}$ for any MPS with LRLE. 

For $D = 3$ we consider first the case of $\S = \sp\{V,W,W^\dg\}$. We define $\tau = \sp\{|x), \ |y)\}$, whereas tracelessness of $W$ and $V$ implies $(x|y) = 0$ and $(x|x) = (y|y)$, respectively (the latter will be taken equal to 1 in the following). It follows that $\S \subset \tau \times \tau$. If w.l.o.g. we set $|x) = |1)$ and $|y) = |2)$, we observe that we retrieve the case of $D = 2$, as 
\begin{align}
A_i^{D=3} = \left(\begin{array}{cc}
A_i^{D=2}&0\\
B_i&c_i
\end{array}\right), \label{3D_2D}
\end{align}
where $A_i^{D=2}$ is proportional to a unitary, and $B_i \in \mathbb{C}^{1 \times 2}$ and $c_i \in \mathbb{C}$ are arbitrary.
If $\S$ is of larger dimension, there are other matrices $V' = |x')(x'| - |y')(y'|$, $W' = |x')(y'|$ and $W'^\dg$ (e.g., $V' = \oe_i (V)$ with $|x') = A_i^\dg |x), \ |y') = A_i^\dg |y)$ etc.), which do not span the same space as $V, \ W$ and $W^\dg$. We define $\tau' = \sp\{|x'), |y')\}$ and take $|x')$ and $|y')$ as orthonormal (which is possible due to $\tr(V') = \tr(W') = 0$). Since $\tau$ and $\tau'$ intersect, we can assume the intersection to be w.l.o.g. along $|y)$.  As $V', \ W'$ and $W'^\dg$ span the full space of traceless matrices contained in $\tau' \times \tau'$, we are allowed to rotate the basis $\{|x'), \ |y')\}$ for $\tau'$ such that $|y) = |y')$.  
Then $\sp\{V, \ W, \ W^\dg\}$ and $\sp\{V', \ W', \ W'^\dg\}$ might only intersect in $|y)(y|$, which is, however, not traceless. Consequently, $\S$ is at least of dimension 6.  In the case of $\S = \sp\{V, \ W, \ W^\dg, \ V', \ W', W'^\dg\}$ it is simple to construct the orthogonal complement of $\S$, which is 
$\overline{\S} = \sp\{\mathbb{I}, |n)(n'|, |n')(n|\}$, where $|n)$ and $|n')$ are the normal vectors of $\tau$ and $\tau'$, respectively. This structure of $\overline{\S}$ uniquely defines $\tau$ and $\tau'$ and vice versa. Therefore, another $\tau'' \neq \tau, \ \tau'$ would not be consistent with this $\overline{\S}$, and the only possibility would be $\overline{\S} = \sp\{\mathbb{I}\}$ corresponding to $A_i A_i^\dg \propto \mathbb{I}$ (i.e., $\mathrm{dim}(\S) = 8$). 
Thus, $\mathrm{dim}(\S) = 6$ is the only remaining case which might lead to a non-trivial MPS. In this case there exist the two subspaces $\tau$ and $\tau'$ of the afore-mentioned type. Based on their properties an ansatz for $A_i$ is made in Appendix \ref{A_i_cases}, which after an elaborate case differentiation leads to 
\begin{align}
A_i \propto [e^{i \phi_i} |1) + e^{i \phi_i'} |3)](l| + e^{i \phi_i''} |2)(m|, 
\end{align}
with $l, m = 1, \ 2, \ 3$, $l \neq m$, as in Example 2.



Within the framework of MPS we have specified the localizable entanglement of a spin chain with one auxiliary spin at each of the borders. We have shown that LRLE can be detected by placing qubits at the ends of the chain. Based on that we were able to derive a theorem according to which it can be checked directly from the matrices of the MPS, whether it possesses LRLE. How this can been done in practice has been indicated by eqs. \eqref{Groebner_first} - \eqref{Groebner_last}, which is a polynomial set of equations that can be solved numerically in a systematic manner. Furthermore,  we provided non-trivial examples of MPS, for which those equations have a simultaneous solution, determining the full sets of MPS with LRLE for $D = 2$ and $D = 3$. 


This work was supported by the Spanish grants S2009/ESP-1594, MTM2011-26912, the European projects QUEVADIS and AQUTE and the CHIST-ERA project CQC.
We are grateful to the Benasque Center for Science and the Perimeter Institute for their hospitality during the course of this work. T. B. W. would like to thank T. S. Cubitt for helpful suggestions. 

\appendix

\section{Proof that LRLE for $D' > 2$ implies LRLE for $D' = 2$}\label{app_show_qubits}

We would like to show that if the LE acquires a finite value for some $D' > 2$, isometries $\P, \ \mathbb{Q} \in \mathcal{P} := \{\P: \mathbb{C}^2 \rightarrow \mathbb{C}^D \ \mathrm{s. t.} \ \P^\dg \P = \mathbb{I}_{2 \times 2} \}$ can be chosen such that \eqref{LE_det}
is non-vanishing in the thermodynamic limit.

We define the coefficient matrix $\Psi_{\mb i}$ of $| \psi^\mathcal{M}_{\mb i} \rangle$ via $(\Psi_{\mb i})_{kl} = \frac{1}{\sqrt{p_{\mb i}}} (k|\P^\dg A_{i_1} ... A_{i_N} \mathbb{Q}|l)$ (here $\P, \ \mathbb{Q}: \mathbb{C}^{D'} \rightarrow \mathbb{C}^D$ can be arbitrarily chosen). Let $\tilde{\P}_{\mb i}$ and $\tilde{\mathbb{Q}}_{\mb i}$ denote the isometries contained in $\mathcal{P}$ projecting $A_{i_1} ... A_{i_N}$ into the subspaces of dimension 2 corresponding to the maximum Schmidt coefficients. For the resulting state we obtain the coefficient matrix $\tilde{\Psi}_{\mb i}= \frac{1}{\sqrt{\tilde{p}_{\mb i}}} \tilde{\P}^\dg_{\mb i} A_{i_1} ... A_{i_N} \tilde{\mathbb{Q}}_{\mb i}$, where $\tilde p_{\mb i}$ is the probability which arises from keeping track only of the projection into the space of the two highest Schmidt coefficients, i.e., $\tilde p_{\mb i} = \tr(\tilde{\P}^\dg_{\mb i} A_{i_1} ... A_{i_N} \tilde{\mathbb{Q}}_{\mb i} \tilde{\mathbb{Q}}_{\mb i}^\dg A_{i_N}^\dg ... A_{i_1}^\dg \tilde{\P}_{\mb i})$. Now, we want to show that
\begin{align}
p_{\mb i} E(\Psi_{\mb i}) \leq f(D') \tilde p_{\mb i} C(\tilde{\Psi}_{\mb i}),
\label{inequ}
\end{align}
where $E(\cdot)$ is the entropy of entanglement \cite{EoE}, $C(\cdot)$ the concurrence \cite{concurrence} and $f(D')$ some finite-valued function to be calculated below. $E(\cdot)$ is given by the von Neumann entropy of the normalized reduced density operator of the bipartite state $\rho_{AB}$, $E(\rho_{AB}) = - \tr(\rho_A \log_2(\rho_A))$ with $\rho_A = \tr_B(\rho_{AB})$.
The concurrence, $C(\cdot)$, is defined for a pure state $| \psi \rangle$ of two qubits as $C(|\psi\rangle) = | \langle \psi^* | \sigma_y \otimes \sigma_y | \psi \rangle | = 2 |\det (\Psi)| \leq 1$, where $\Psi$ is the $2 \times 2$ coefficient matrix of the state.

We drop the index $\mb i$ for the variables introduced in the following and denote by $Q_1 \geq Q_2 \geq ... \geq Q_{D'}$ the eigenvalues of the unnormalized 1-particle reduced density operator, $\rho'_A$, of $\rho'_{AB} = p_{\mb i} |\psi_{\mb i}^\mathcal{M}\rangle \langle \psi_{\mb i}^\mathcal{M}| = p_{\mb i} \, \rho_{AB}$ after the measurement. Thus, we get
\begin{align}
p_{\mb i} = \sum_{k=1}^{D'} Q_k \leq (Q_1 + Q_2) \frac{D'}{2} = \tilde{p}_{\mb i} \frac{D'}{2},
\end{align}
and it is sufficient to show that
\begin{align}
E(\Psi_{\mb i}) \leq g(D') C(\tilde{\Psi}_{\mb i}). \label{show_g}
\end{align}
In order to do so we first derive an upper bound for the LHS of \eqref{show_g}. The eigenvalues of the normalized 1-particle reduced density operator, $\rho_A$, of $\rho_{AB} = |\psi_{\mb i}^{\mathcal{M}} \rangle \langle \psi_{\mb i}^\mathcal{M} |$ are $P_k = {Q_k}/{\sum_n Q_n}$. Hence, we obtain
\begin{align}
E(\Psi_{\mb i}) &= E(\rho_{AB}) = - \sum_{k=1}^{D'} P_k \log_2 P_k \notag \\
&\leq -P_1 \log_2 P_1 - (1-P_1) \log_2 \frac{1-P_1}{D'-1} \notag \\
&:= w(P_1, D'),
\end{align}
since the sum from the second term on is maximized for $P_2 = ... = P_{D'} = (1-P_1)/(D'-1)$. 

A lower bound for the RHS of \eqref{show_g} can be found by using the entropy of entanglement of the qubit system into which has been projected, $E(\tilde{\Psi}_{\mb i}) = \mathcal{F}(C(\tilde{\Psi}_{\mb i})) \leq C(\tilde{\Psi}_{\mb i})$ \cite{concurrence}, where $\mathcal{F}(C) = h ((1+ \sqrt{1-C^2})/2)$ and $h(x) = -x \log_2 x - (1-x) \log_2 (1-x)$, and thus $\mathcal{F}(C) \leq C$ for $0 \leq C \leq 1$. In our case by definition of $\tilde{\P}_{\mb i}$ and $\tilde{\mathbb{Q}}_{\mb i}$ the entropy of entanglement is
\begin{align}
E(\tilde \Psi_{\mb i}) &= - \frac{P_1}{P_1 + P_2} \log_2 \frac{P_1}{P_1 + P_2} - \frac{P_2}{P_1 + P_2} \log_2 \frac{P_2}{P_1 + P_2} \notag \\
&\geq - \frac{P_1 (D' - 1)}{P_1(D'-1) + 1 - P_1} \log_2 \frac{P_1 (D'-1)}{P_1 (D' - 1) + 1 - P_1} \notag \\
&- \frac{1-P_1}{P_1(D'-1) + 1 - P_1} \log_2 \frac{1-P_1}{P_1(D'-1) + 1 - P_1}, \notag \\
&:= r(P_1, D')
\end{align}
since for fixed $P_1$ the uppermost sum is minimized for $P_2 = \frac{1-P_1}{D'-1}$. Therefore, we have
\begin{align}
\frac{E(\Psi_{\mb i})}{C(\tilde{\Psi}_{\mb i})} \leq \frac{w(P_1, D')}{r(P_1, D')},
\end{align}
which is finite for $\frac{1}{D'} \leq P_1 < 1$. In the limit $P_1 \rightarrow 1$ the ratio takes the value $D' - 1$, thus verifying that there exists a finite-valued function $g(D')$ such that ${E(\Psi_{\mb i})}/{C(\tilde{\Psi}_{\mb i})} \leq g(D')$. 

Now we apply \eqref{inequ} to bound the LE for arbitrary $D'$
\begin{align}
L(\rho) = \sup_{\{|i\rangle\}} \sum_{\mb i} p_{\mb i} E(\Psi_{\mb i}) \leq f(D') \sup_{\{|i\rangle\}} \sum_{\mb i} \tilde{p}_{\mb i} C(\tilde{\Psi}_{\mb i})
\end{align}
with $\tilde{\Psi}_{\mb i}= \frac{1}{\sqrt{\tilde{p}_{\mb i}}} \tilde{\P}^\dg_{\mb i} A_{i_1} ... A_{i_N} \tilde{\mathbb{Q}}_{\mb i}$. In analogy to \eqref{LE_det} we obtain
\begin{align}
L(\rho) \leq 2 f(D') \sup_{\{|i\rangle\}} \sum_{i_1, ..., i_N} |\det(\tilde{\P}_{\mb i}^\dg A_{i_1} ... A_{i_N} \tilde{\mathbb{Q}}_{\mb i})|.
\end{align}
We introduce an $\epsilon$-net ($\epsilon > 0$) $\mathcal{N}_{\epsilon}$ of isometries contained in $\mathcal{P}$, that is, for any such isometry $\tilde{\P}$ there exists a $\mathbb{V} \in \mathcal{N}_{\epsilon}$ with operator norm
$\| \tilde{\P} - \mathbb{V} \| \leq \epsilon$. By choosing $\epsilon$ sufficiently small we get up to orders in $\epsilon$
\begin{align}
L(\rho) &\leq 2 f(D') \sum_{\mathbb{V} \in \mathcal{N}_\epsilon} \sum_{\mathbb{W} \in \mathcal{N}_\epsilon} \sum_{i_1, ..., i_N} | \det(\mathbb{V}^\dg A_{i_1} ... A_{i_N} \mathbb{W})| \notag \\
&\times \delta_{\mathbb{V}, \tilde{\P}_{\mb i}} \delta_{\mathbb{W}, \tilde{\mathbb{Q}}_{\mb i}} \\
&:= 2 f(D') \sum_{\mathbb{V} \in \mathcal{N}_\epsilon} \sum_{\mathbb{W} \in \mathcal{N}_\epsilon} \omega(\mathbb{V}, \mathbb{W}) \label{omega}.
\end{align}

If the LE is non-zero in the thermodynamic limit, \eqref{omega} is lower bounded by $L(\rho) > 0$. Consequently, for all $r \gg \epsilon$ there must exist $r$-regions $\mathcal{R}^1_r, \ \mathcal{R}^2_r \subseteq \mathcal{N}_{\epsilon}$ containing isometries $\mathbb{V}', \ \mathbb{W}'$ such that $\forall \ \mathbb{V} \in \mathcal{R}^1_r$ and $\forall \ \mathbb{W} \in \mathcal{R}^2_r$ we have $\|\mathbb{V} - \mathbb{V}'\| \leq r$ and $\|\mathbb{W} - \mathbb{W}'\| \leq r$, respectively, with
\begin{align}
2 f(D') \sum_{\mathbb{V} \in \mathcal{R}_r^1} \sum_{\mathbb{W} \in \mathcal{R}_{r}^2} \omega(\mathbb{V}, \mathbb{W}) \geq L(\rho) \frac{V_{\mathcal{R}^1_r} V_{\mathcal{R}^2_r}}{V_{\mathcal{N}_\epsilon}^2},
\label{volume}
\end{align}
where $V_{\mathcal{R}}$ denotes the volume, i.e., the number of points in the $\epsilon$-discretization of region $\mathcal{R}$. By choosing $r$ small enough, one obtains $\omega(\mathbb{V}, \mathbb{W}) = \omega(\mathbb{V}', \mathbb{W}')(1+\mathcal{O}(r))$, such that sum in the LHS of \eqref{volume} contains ${V_{\mathcal{R}^1_r} V_{\mathcal{R}^2_r}}/{V_{\mathcal{N}_\epsilon}^2}$ constant terms, and therefore
\begin{align}
2 f(D') \omega(\mathbb{V}', \mathbb{W}') \left(1 + \mathcal{O}(r)\right) \geq L(\rho).
\end{align}
Hence, for sufficiently small $r$ one gets $\omega(\mathbb{V}', \mathbb{W}') > 0$, i.e., \eqref{LE_det} is non-vanishing for some isometries $\mathbb{V}'$ and $\mathbb{W}'$, if there is LRLE for some $D' > 2$ \qed


\section{Proof that all factors but the last one of \eqref{projectors} have to be 1}\label{Lemma2}

We want to demonstrate that in the limit $N \rightarrow \infty$ the sum \eqref{projectors}
can be non-zero only if one can choose $\P^\dg$ such that for all well-defined $\P^{j-1 \,\dg}_{i_1, ..., i_{j-1}}$ (i.e. those for which $\P^\dg A_{i_1} ... A_{i_{j-1}} \neq 0$) the inequality
\begin{align}
\sum_{i_j} \left| \det (\P^{j-1 \dg}_{i_1,...,i_{j-1}} A_{i_j} \P^j_{i_1,..., i_{j}}) \right| \leq 1
\end{align}
is an equality.

\noindent Formally this claim reads
\begin{align}
\mathrm{LRLE} \ \Rightarrow \ \exists \ \P^\dg \ \mathrm{s.t.}\ \forall \ i_1, \ ..., \ i_{j-1} = 1, \ ...,\ d \ \notag \\
\sum_{i_j} \left| \det\left( \P^{j-1 \,\dg}_{i_1, ..., i_{j-1}} A_{i_j} \P^{j}_{i_1, ..., i_j} \right)\right| = 1,
\end{align}
or equivalently,
\begin{align}
\neg \, \mathrm{LRLE} \ &\Leftarrow \ \forall \ \P^\dg \ \exists \ (i_1, \ ..., \ i_{j-1}) \ \mathrm{s.t.} \notag \\ &\sum_{i_j} \left| \det\left( \P^{j-1 \,\dg}_{i_1, ..., i_{j-1}} A_{i_j} \P^{j}_{i_1, ..., i_j} \right)\right| \neq 1,
\label{lemma2}
\end{align}
which is what we want to show in the following. Thus, we assume that for the sum
\begin{align}
\sum_{i_1, ..., i_s} \left| \det \left( \P^\dg A_{i_1} ... A_{i_s} \P^{s}_{i_1 ... i_s} \right)\right| := 1 - \Delta^{(s)}(\P^\dg)
\label{sum_Delta}
\end{align}
there is a minimum integer $s_*$ such that
\begin{align}
\min_{\P^\dg} \Delta^{(s_*)}(\P^\dg) := \Delta_* > 0.
\label{min_Delta}
\end{align}
After building blocks of $s$ terms in \eqref{projectors}, each of them can be upper bounded by
\begin{align}
1 - \Delta(\P_{i_1, ..., i_{j}}^{j \, \dg}) \leq 1 - \Delta_*,
\label{star}
\end{align}
which shows that the LE must fulfill
\begin{align}
L(\rho) \leq (1 -\Delta_*)^{\frac{N}{s_*}-1} \xrightarrow{N \rightarrow \infty} \ 0,
\end{align}
i.e., there is no LRLE. \qed


\section{Proof that \eqref{basic_rel} is also a sufficient condition}\label{app_suffQ}

Here we show that $\mathbb{Q}$ can always be chosen such that \eqref{projectors} is finite whenever \eqref{basic_rel}
is fulfilled. We denote the proportionality factor in \eqref{basic_rel} by $\gamma_{i_1, ..., i_j}$ for which $\mathcal{E}(\mathbb{I}) = \sum_{i=1}^d A_i A_i^\dg = \mathbb{I}$ implies $\sum_{i_j} | \gamma_{i_1, ..., i_j}|^2 = | \gamma_{i_1, ..., i_{j-1}}|^2$. \eqref{LE_det} thus reads
\begin{align}
L(\rho) = 2 \sum_{i_1, ..., i_N} \left| \det(\P^{N \, \dg}_{i_1 ...i_N} \mathbb{Q})\right| |\gamma_{i_1 ... i_N}|^2.
\end{align}
Following the approach in Part \ref{app_show_qubits} we introduce an $\epsilon$-net $\mathcal{N}_\epsilon$ in the isometries of $\mathcal{P}$ and obtain up to order $\epsilon$
\begin{align}
L(\rho) = 2 \sum_{\mathbb{V} \in \mathcal{N}_\epsilon} \sigma(\mathbb{V}) | \det(\mathbb{V}^\dg \mathbb{Q})|, \label{sum_V}
\end{align}
where we defined
\begin{align}
\sigma(\mathbb{V}) = \sum_{i_1 ... i_N} |\gamma_{i_1 ... i_N}|^2 \delta_{\P^{N}_{i_1 ... i_N}, \mathbb{V}}.
\end{align}
Consider now the sum
\begin{align}
\sum_{\mathbb{V} \in \mathcal{N}_\epsilon} \sigma(\mathbb{V}) = \sum_{i_1 ... i_N} |\gamma_{i_1, ..., i_N}|^2 = 1.
\end{align}
Since it takes a finite value, for any $r \gg \epsilon$ there must exist an $r$-region $\mathcal{R}_r \subseteq \mathcal{N}_\epsilon$ containing $\mathbb{V}''$ such that $\forall \ \mathbb{V} \in \mathcal{R}_r$ the relation $\| \mathbb{V} - \mathbb{V}''\| \leq r$ holds with
\begin{align}
\sum_{\mathbb{V} \in \mathcal{R}_r} \sigma(\mathbb{V}) \geq \frac{V_{\mathcal{R}_r}}{V_{\mathcal{N}_\epsilon}},
\end{align}
where ${V_{\mathcal{R}_r}}$ and ${V_{\mathcal{N}_\epsilon}}$ are the number of points of the respective regions in the $\epsilon$-grid. For sufficiently small $r$ we consider only the following part of the sum in \eqref{sum_V}
\begin{align}
&2 \sum_{\mathbb{V} \in \mathcal{R}_r} \sigma(\mathbb{V}) | \det(\mathbb{V}^\dg \mathbb{Q})| \notag \\
&= 2 \sum_{\mathbb{V} \in \mathcal{R}_r} \sigma(\mathbb{V}) | \det({\mathbb{V}''}^\dg \mathbb{Q})| \left(1 + \mathcal{O}(r)\right).
\end{align}
If we set $\mathbb{Q} = \mathbb{V}''$, this is lower bounded by $2(1 + \mathcal{O}(r)){V_{\mathcal{R}_r}}/{V_{\mathcal{N}_\epsilon}}$, i.e.,
\begin{align}
L(\rho) \geq 2(1 + \mathcal{O}(r))\frac{V_{\mathcal{R}_r}}{V_{\mathcal{N}_\epsilon}},
\end{align}
which is small but positive for sufficiently small $r$. \qed


\section{Matrices with LRLE for $D = 3$ and $\dim(\S) = 6$}\label{A_i_cases}

For $D = 3$ in the main text it has been shown that the case of $\dim(\S) = 3$ corresponds to a trivial extension of the matrices for $D = 2$, $A_i^{D=2} = \alpha_i U_i$, to $D = 3$. Moreover, it has been noted that larger possible dimensions of $\S$ are only 6 and 8, the latter implying $A_i A_i^\dg \propto \mathbb{I}$. For the case of $\dim(\S) = 6$ it has been shown that 
$\S = \sp\{V, \ W, \ W^\dg, \ V', \ W', W'^\dg\}$, where $V = |x)(x| - |y)(y|$, $W = |x)(y|$ and analogous definitions with $|x')$ and $|y')$. The reason for the introduction of $V'$ and $W'$ of the same type as $V$ and $W$ is that $\S$ is closed under any $\oe_i$, which maps, e.g., $V$ to $V_i = A_i^\dg |x)(x|A_i - A_i^\dg |y)(y|A_i$. For some $\oe_i$ 
$V_i$ must happen not to be in $\sp\{V, \ W, W^\dg\}$ if $\dim(\S) > 3$. We see that in this case $A_i^\dg|x)$ and $A_i^\dg |y)$ are new orthogonal vectors with the same norm (due to $\tr(V_i) = \tr(W_i) = 0$). From them we obtain the orthonormal vectors $|x') = [(x|A_i A_i^\dg |x)]^{-1/2} A_i^\dg |x)$ and $|y') = [(y|A_i A_i^\dg |y)]^{-1/2} A_i^\dg |y)$. Thus, $A_i$ applied from the right maps all vectors contained in $\tau = \sp\{|x), \ |y)\}$ to vectors in $\tau' = \sp\{|x'), \ |y')\}$, while preserving their relative lengths, or equivalently, the angles between them. 

As noted in the main text, the existence of another $\tau'' \neq \tau, \ \tau'$ (supporting $V'', \ W''$ and $W''^\dg$) is excluded for $\dim(\S) = 6$. Consequently, any 
$A_i$ applied from the right has to map $\tau$ to $\tau$ or $\tau'$ and $\tau'$ to $\tau$ or $\tau'$ in such a way that all angles between vectors lying in one of those subspaces are preserved. $\tau$ and $\tau' \neq \tau$ at this point can be arbitrary two-dimensional linear subspaces of $\mathbb{C}^3$, whereas we assume their intersection to be along $|y) = |y')$. The justification for this choice is that for any orthonormal pair $|x'), \ |y')$ of vectors in $\tau'$ $\sp\{V', \ W', \ W'^\dg\}$ is the full subspace of traceless matrices contained in $\tau' \times \tau'$. After choosing w.l.o.g. $|x) := |1)$ and $|y) := |2)$, we can make the ansatz
\begin{align}
A_i = \gamma_i \left(\begin{array}{ccc}
1&0&0\\
0&1&0\\
a_i&b_i&c_i
\end{array}
\right)U_i,
\end{align}
$\gamma_i, \ a_i, \ b_i, \ c_i \in \mathbb{C}$ and $U_i$ is a unitary. We see that indeed $\tr(A_i^\dg V A_i) = 0$, i.e., $(1|A_i A_i^\dg |1) = (2| A_i A_i^\dg |2) = |\gamma_i|^2$ and $\tr(A_i^\dg W A_i) = 0$, i.e., $(2|A_i A_i^\dg |1) = 0$ are fulfilled. 
The normal vector of $\tau$ is $|n) = |3)$, and the one of $\tau'$ is of the form $|n') = r |1) + s|3)$, since $\tau$ and $\tau'$ intersect in the $|2)$-axis. Because of $\tr(A_i^\dg S A_i) = 0 \ \forall \ S \in \S$, it follows that $A_i A_i^\dg \in \overline{\S} = \sp\{\mathbb{I}, |n)(n'|, |n')(n|\}$. Therefore, 
\begin{align}
A_i A_i^\dg = \alpha_i \mathbb{I} + \beta_i \big(|3)[r(1| + s(3|] + [r^*|1) + s^* |3)](3|\big),
\end{align}
which results in $b_i = 0 \ \forall \ i = 1, \ ..., \ d$. Now, the matrix
\begin{align}
M_i = \left(\begin{array}{ccc}
1&0&0\\
0&1&0\\
a_i&0&c_i
\end{array}\right)
\end{align}
applied from the right maps $\tau$ to $\tau$ and $\tau'$ to, say, $\tilde \tau$ preserving the angles between vectors lying in one of them. As a result, there are four possible cases for the action of $U_i$
\begin{align}
a) \ \ U_i:& \ (\tau, \ \tilde \tau) \rightarrow (\tau, \ \tau), \\
b) \ \ U_i:& \ (\tau, \ \tilde \tau) \rightarrow (\tau, \ \tau'), \\
c) \ \ U_i:& \ (\tau, \ \tilde \tau) \rightarrow (\tau', \ \tau), \\
d) \ \ U_i:& \ (\tau, \ \tilde \tau) \rightarrow (\tau', \ \tau'). 
\end{align}
a) and d) imply $\tilde \tau = \tau$ and hence $a_i = e^{i \varphi}$ and $c_i = 0$, whereas b) and c) mean $\tilde \tau \neq \tau$. Therefore, in the latter case the intersection of $\tau$ and $\tilde \tau$ is the $|2)$-axis. Since it has to be mapped by $U_i$ to the intersection of $\tau$ and $\tau'$, which is likewise the $|2)$-axis, $|2)$ must be an eigenvector of $U_i$. Then, it follows that $U_i$ has a unitary action in the $|1)$-$|3)$-plane. Hence, b) corresponds to $\tilde \tau = \tau'$, and therefore $a_i = 0$, \ $c_i = e^{i \varphi}$. Last, in c) double application of $U_i$ would map $\tilde \tau \xrightarrow{U_i} \tau \xrightarrow{U_i} \tau'$, wherefore $M_i$ has to carry out a reflection of $\tau'$ on the $\tau$-plane, i.e., $a_i = 0$ and $c_i = -1$. 
We thus obtain either $a_i = 0$ and $c_i = e^{i \varphi}$ or $a_i = e^{i \varphi}$ and $c_i = 0 \ \forall \ i = 1, \ ..., \ d$ and conclude that $\tau' = \sp\{|2), |3)\}$.

The requirement that $A_i$ maps $\tau$ to $\tau$ or $\tau'$ and $\tau'$ to $\tau$ or $\tau'$ also holds for $A_i^2$, from which we deduce that $U_i$ must contain only one non-vanishing entry per row and column (with magnitude 1). All results combined together yield
\begin{align}
A_i \propto [e^{i \phi_i} |1) + e^{i \phi_i'} |3)](l| + e^{i \phi_i''} |2)(m|, 
\end{align}
with $l, m = 1, \ 2, \ 3$, $l \neq m$.


\begin{thebibliography}{99}

\bibitem{LE} F. Verstraete, M. Popp, and J. I. Cirac, Phys. Rev. Lett. \textbf{92}, 027901 (2004).
\bibitem{repeaters} H.-J. Briegel, W. D\"{u}r, J. I. Cirac, and P. Zoller, Phys. Rev. Lett. \textbf{81}, 5932 (1998).
\bibitem{LEMPS} F. Verstraete, M. A. Martin-Delgado, and J. I. Cirac, Phys. Rev. Lett. \textbf{92}, 087201 (2004).
\bibitem{LECondMat} R. Orus, and H.-H. Tu, Phys. Rev. B \textbf{83}, 201101(R) (2011); S. O. Skr{\o}vseth, and S. D. Bartlett, Phys. Rev. A \textbf{80}, 022316 (2009);
J. K. Pachos, and M. B. Plenio, Phys. Rev. Lett. \textbf{93}, 056402 (2004); 
B.-Q. Jin, and V. E. Korepin, Phys. Rev. A \textbf{69}, 062314 (2004).
\bibitem{Gaussian} J. Fiur\'{a}\v{s}ek, and L. Mi\v{s}ta, Jr., Phys. Rev. A \textbf{75}, 060302(R) (2007); A. Serafini, G. Adesso, and F. Illuminati, Phys. Rev. A \textbf{71}, 032349 (2005); G. Adesso, and F. Illuminati, Phys. Rev. Lett. \textbf{95}, 150503 (2005).
\bibitem{Ent_trans} V. Subrahmanyama, and A. Lakshminarayanb, Phys. Lett. A \textbf{349}, 164 (2006).
\bibitem{Sahoo} L. Campos Venuti, and M. Roncaglia, Phys. Rev. Lett. \textbf{94}, 207207 (2005); S. Sahoo, arXiv:1201.5620v4.
\bibitem{Ent_Length} The case of LRLE needs to be differentiated from the one of diverging entanglement length introduced in \cite{LE}: A divergence of the entanglement length only implies that the LE cannot decay exponentially with the spin distance, however it still may decay algebraically, c.f. \cite{LE_long}. 
\bibitem{MPS} D. P\'{e}rez-Garc\'{i}a, F. Verstraete, M. M. Wolf, and J. I. Cirac, Quant. Inf. Comp. \textbf{7}, 401 (2007).
\bibitem{SchoenPRL} C. Sch\"{o}n, E. Solano, F. Verstraete, J. I. Cirac, and M. M. Wolf, Phys. Rev. Lett. \textbf{95}, 110503 (2005).
\bibitem{FNW92} M. Fannes, B. Nachtergaele, R. F. Werner, Commun. Math. Phys. \textbf{144}, 443 (1992).
\bibitem{LE_long} M. Popp, F. Verstraete, M. A. Martin-Delgado, and J. I. Cirac, Phys. Rev. A \textbf{71}, 042306 (2005).
\bibitem{QMC} O. F. Sylju\r{a}sen, Phys. Lett. A \textbf{322}, 25 (2004);
M. Popp, F. Verstraete, M. A. Martin-Delgado, and J. I. Cirac, Appl. Phys. B \textbf{82}, 225 (2006);
P. Androvitsaneas, E. Paspalakis, and A. F. Terzis, Ann. Phys. \textbf{327}, 212 (2012).
\bibitem{concurrence} W. K. Wootters, Phys. Rev. Lett. \textbf{80}, 2245 (1998).
\bibitem{AKLT} F. Verstraete, M. A. Martin-Delgado, and J. I. Cirac, Phys. Rev. Lett. \textbf{92}, 087201 (2004).
\bibitem{D_eq2} Note that for $D = 2$ all isometries are unitaries, i.e., according to \eqref{new_isometry} all matrices $A_i$ have to be proportional to unitaries themselves. 
\bibitem{Groebner} H. Hironaka, Ann. Math. \textbf{79}, 109 (1964); B. Buchberger, Austria, Universit\"{a}t Innsbruck, Diss., 1965.
\bibitem{EoE} C. H. Bennett, H. J. Bernstein, S. Popescu, and B. Schumacher, Phys. Rev. A \textbf{53}, 2046 (1996).


\end{thebibliography}

\end{document}